\documentclass{article}
\usepackage{epsf}
\def\RN{Reis\-sner-Nord\-str\"{o}m }
\begin{document}
\begin{flushright}
Alberta-Thy-35-94 \\
gr-qc/9411050
\end{flushright}
\begin{center}
\vfill
{\huge \bf Structure of the Spherical Black Hole Interior}
\vfill
{\LARGE \bf  A. Bonanno\footnote[2]{Permanent address: Institute of Astronomy,
University of Catania, Viale Andrea Doria 6, 95125 Catania Italy}, S. Droz, W.
Israel and  S.M. Morsink
\vspace{1cm}}\\
{\large
Canadian Institute for Advanced Research Cosmology Program,\\
Theoretical Physics Institute, University of  Alberta,\\
Edmonton, Alberta, Canada T6G 2J1}

\vfill
{\Large \bf
Abstract}
\end{center}

\begin{quote}
{\large
\baselineskip 14pt
The internal structure of a charged spherical black hole is still a topic of
debate.
In a nonrotating but aspherical gravitational collapse to form a spherical
charged black hole, the backscattered gravitational wave tails enter the black
hole  and are blueshifted at the Cauchy horizon. This has a catastrophic effect
if combined with an outflux crossing the Cauchy horizon: a singularity develops
at the Cauchy horizon and the effective mass inflates. Recently a numerical
study of a massless scalar field in the \RN background suggested that a
spacelike singularity may form before the Cauchy horizon forms. We will show
that there exists an approximate analytic solution of the scalar field
equations which allows the mass inflation singularity at the Cauchy horizon to
exist.  In particular, we see no evidence that the Cauchy horizon is preceded
by a  spacelike singularity.}
\end{quote}\vfill
\normalsize
\twocolumn
\section{Introduction}

The final state of a star that undergoes gravitational collapse into a black
hole is described by the uniqueness theorems of general relativity \cite{Wald}.
At late times, after the star's irregularities have been radiated away
 \cite{Price1}  the external geometry is described by the stationary
Kerr-Newman
solution. But the structure of the realistic black hole's interior has not been
definitively determined and is still the subject of much debate.

The unproven, yet plausible strong cosmic censorship principle   leads one to
suspect that the singularity in a physical black hole ought to be spacelike,
and described by the general mixmaster type solution  \cite{BKL}. But the
Kerr-Newman singularity at $r=0$ is timelike. How could the inclusion of the
details of collapse alter this picture so drastically? Penrose has described
the basic mechanism \cite{Penrose}. In the Kerr-Newman solution (and the \RN
solution), the timelike singularity is preceded by a Cauchy horizon. The Cauchy
horizon is the boundary of predictability, making the existence of a timelike
singularity in a region beyond it physically irrelevant.
The Cauchy horizon is  also a surface of infinite blueshift. Any freefalling
observer crossing the Cauchy horizon will measure infalling radiation to have
infinite energy density. When only ingoing radiation is present a weak
nonscalar singularity forms, which is classified as a whimper
singulartity \cite{Ellis}. Whimper singularities are known to be unstable to
perturbations which can transform them into  stronger
singularities \cite{Ellis}.

This scenario can be examined more closely by assuming spherical symmetry.
When an outflow crossing the Cauchy horizon is included in the analysis of a
charged spherical black hole  \cite{Poisson,Ori}, the mass function is found to
diverge exponentially. In spherical symmetry, the only non-vanishing component
of the Weyl  tensor, the  invariant $\Psi_2$, is proportional to the mass
function. The net result of including an outflux is that a scalar curvature
singularity forms along the Cauchy horizon. This phenomenon has been dubbed
``mass inflation" \cite{Poisson}.The \RN black hole has the same causal
structure as the Kerr-Newman black hole, so we expect that results for
spherical symmetry should be qualitatively the same for the more general black
hole.
Preliminary calculations suggest that the general picture derived for
spherical symmetry does not change dramatically in a non-spherical black
hole \cite{Ori2}.

The important issue, which will be addressed in this paper, is to ask how
generic is the mass inflation picture.  General stability
arguments \cite{Yurtsever} and the numerical study of a scalar
field \cite{Gnedin} suggest that the Cauchy horizon is preceded by a spacelike
$r=0$ singularity. Hence evolution of the interior should end before the Cauchy
horizon forms. But while these studies are suggestive, a conclusive analysis of
the internal structure is still lacking even for spherical holes.

\epsfxsize=7.5cm
\epsffile{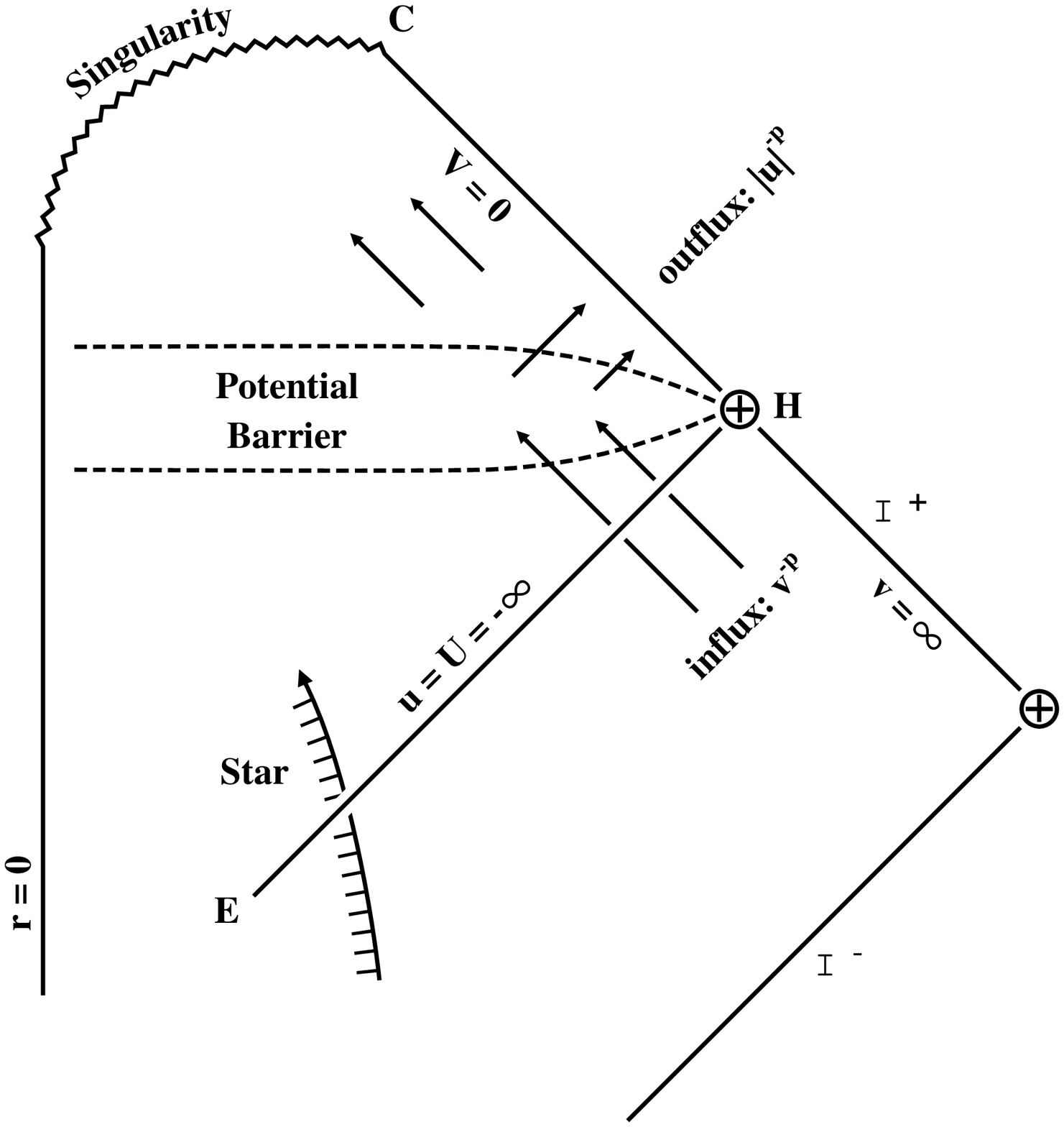}
\begin{center}
\parbox{6.5cm}{
\small
Fig 1. Penrose conformal diagram of a collapsing star. Note that the point H
is not part of the manifold, but a singular point of this mapping.}
\end{center}
\mbox{\hspace{1cm}}

Our main objective is to examine the ``corner" region, (``point" H in Fig. 1),
bounded by the Cauchy and event horizons, and to decide whether it is possible
for an early portion of the Cauchy horizon to survive. We have found, using
analytic approximations and physical arguments, that there is a general
solution of the scalar field equations which is consistent with the mass
inflation picture \cite{Bonanno}.

The description of the black hole interior is simplified by causality: behind
the event horizon the coordinate $r$ is timelike, so a descent into a black
hole is an evolution in time. The solution down to any particular radius is
only influenced by the initial data at larger radii.  Most importantly, our
lack of understanding of the true quantum description of gravity (applicable in
the innermost layers) will not affect the description of the outer classical
and semi-classical layers.  The classical evolution problem is thus on firm
ground.

  In an aspherical collapse with zero angular momentum, the asphericities must
be radiated away in order to form a static spherically symmetric black hole, as
the no hair theorems demand.  Price \cite{Price1} has shown that the star
settles into an asymptotically  \RN state with perturbations that die off with
advanced time as an inverse power law.  These perturbations are scattered by
the black hole's external potential barrier so that part of the radiation falls
into the black hole. Near the event horizon, the backscattered energy flux
takes
the form $v^{-p}$ where  $v\rightarrow \infty$ corresponds to  $\cal{I^+}$ and
$p=4l+4$, where $l$ is the spherical harmonic multipole of the
perturbation \cite{Gundlach1}.
This result was recently  verified by  numerical studies of the coupled
Einstein-scalar field equations \cite{Gundlach2}. We will use the Price tail
$v^{-p}$ as the initial data on the event horizon, the natural choice for the
initial data surface for the black hole interior evolutionary problem.

Our paper is organised as follows.  In   \S 2  and \S 3 we will review
the mass inflation model of the black hole interior and discuss its
limitations. The boundary conditions will be discussed in section \S 4. These
results will be applied to a null crossflow model in section \S 5. In
\S 6, we present  an  approximate analytic solution to the scalar field
equations inside a spherical black hole.

\section{Schematic Mass Inflation Mo\-dels}
The essential physics of mass inflation -- a blue\-shif\-ted influx combined
with an
outflux -- can be illustrated using simple solvable models. The simplest model
is that of two concentric spherical null shells, one ingoing, the other
outgoing, which cross without interaction. This situation was examined for
shells crossing in the
Schwarzschild  geometry \cite{DTR} and extended to \RN \cite{Barrabes}.

Imagine the spacetime split into four regions $A$, $B$, $C$ and $D$ by the two
crossing shells (see Fig. 1 of   \cite{Barrabes}).  Each region $i$,  is
described by a \RN solution with metric function $f_i = 1 - 2m_i/r + e^2/r^2$.
The masses of the ingoing and outgoing shells (before collision) are $m_{in}
= m_D - m_B$ and $m_{out} = m_C - m_B$. The final result of the
analysis is that the metric functions are related by the generalised DTR
relation \cite{DTR,Barrabes}
\begin{equation}
f_A f_B = f_C f_D,
\end{equation}
which holds at $r_0$, the radius of crossing.
Mass inflation occurs when the ingoing shell is very close (within distance
$\epsilon$) of the Cauchy horizon corresponding to the extension of sector $B$,
the space between them before the shells cross.  Then $f_B \sim \epsilon$,
while the
product $f_C f_D$ stays finite. This forces the metric function in region $A$,
after the shells cross,  to diverge as $f_A \sim 1/\epsilon$.

The crucial role of the outflux is illustrated by writing the expression for
the mass after the crossing,  $m_A$ \cite{Barrabes}
\begin{equation}
m_A = {m_C m_D \over m_B} + { m_{in} m_{out} \over m_B (-f_B)}.
\end{equation}
The first term is finite. When the ingoing shell is close to the Cauchy
horizon, the second term becomes very large. But note that the second term is
proportional to the product $ m_{in} m_{out}$. The absence of the outgoing
shell would render the second term harmless. Thus the outflow, however weak, is
crucial for the divergence of the mass function after the shells cross.

Ori has introduced a generalization of this simple model \cite{Ori}.  He
considers a continuous influx and treats the outflux as a thin shell. This
allows the matching of two ingoing Vaidya solutions along the outgoing
lightlike shell  $\Sigma$, a finite Kruskal time after the event horizon (see
Fig.
1 of  \cite{Anderson}):
\begin{eqnarray}
ds^2 = dv_\pm(f_\pm dv_\pm - 2 dr) + r^2 d\Omega^2,   \label{3} \\
f_\pm = 1 - 2m_\pm(v_\pm)/r + e^2/r^2, \nonumber
\end{eqnarray}
where the subscript + (-) refers to the region after (before) the shell.
The Einstein equations link the mass with the influx
\begin{equation}\label{4}
 L_\pm = dm_\pm / dv_\pm , \
T_{ab}^{\pm} ={L_{\pm}(v_\pm) \over 4 \pi r^2} \partial_a v_\pm \partial_b
v_\pm.
\end{equation}
Continuity of the line element and the radial coordinate, $r$,  yields the
equations
\begin{equation}\label{5}
f_+ dv_+ = f_- dv_- = 2 dr
\end{equation}
along the shell. Continuity of the influx across the shell gives the equation
\begin{equation}\label{6}
{1\over f_+^2} {dm_+ \over dv_+} = {1\over f_-^2} {dm_- \over dv_-} .
\end{equation}
These two equations yield the simple equation  \cite{Anderson}
\begin{equation}\label{7}
{dm_+ \over f_+} = {dm_- \over f_-}
\end{equation}
in which mass inflation is evident. The metric function $f_-$ goes to zero as
the Cauchy horizon is approached, causing the right hand side of the equation
to diverge. The presence of the outgoing shell displaces the apparent horizon
to smaller radii so that $f_+ \neq 0$ at the Cauchy horizon. This equation
implies that beyond the
shell, the mass will diverge at CH.

This model can be solved asymptotically close to the Cauchy horizon. The mass
function prior to the shell must reproduce the Price power law tail: $m_- = m_0
- {\beta \over \kappa_0 (p-1)}(\kappa_0 v_-)^{-(p-1)}$.  Here $v_-$ is the
usual Eddington-Finkelstein advanced time coordinate which is infinite on the
Cauchy horizon, $\beta$ is a dimensionless constant and $\kappa_0$ is the
surface gravity of the inner horizon.
Equation (\ref{5})  is then integrated for $r$ along $\Sigma$ as
$v_-\rightarrow \infty$
\begin{eqnarray}\label{8}
r_\Sigma(v_-) & = &r_0 + {\beta \over r_0 \kappa_0^2 (p-1)} \\
& & (\kappa_0
v_-)^{-(p-1)} \left(1 + {(p-1) \over \kappa_0 v_-} + ...\right). \nonumber
\end{eqnarray}
Equation (\ref{7}) can now  be integrated using (\ref{8}) to show that the mass
function diverges exponentially
\begin{equation}
m_+(v_-) \sim e^{\kappa_0 v_-} (\kappa_0 v_-)^{-p} ,\  v_- \rightarrow \infty
\  .
\end{equation}
This phenomenon has been dubbed mass inflation \cite{Poisson}. Indeed this is a
scalar curvature singularity since
the   Weyl curvature  invariant diverges exponentially, $\Psi_2 \sim
e^{\kappa_0 v_-}/r_0^2$ as the Cauchy horizon is approached.

%
%
 It is worth noting that there are coordinates in which the metric is
finite. Near the CH the line element  is given to a very good approximation
by
\begin{equation}
ds^2 = 2 \frac{dv_+}{r} \left(r dr + m_+(v_+) dv_+ \right) + r^2
d\Omega^2
    \label{10}
\end{equation}
where $v_+$ is the standard advanced time coordinate {\em behind}
$\Sigma$. It
is easily checked, that the coordinate $u$, defined by
\[ du = r dr + m_+(v_+) dv_+ \] is regular at the CH. (\ref{3}) now
becomes
\[
ds^2 = 2 \frac{dv_+ du}{r} + r^2 d\Omega^2.
\]
The mass inflation singularity, though much stronger than a whimper
singularity is still very weak in this sense. It is exactly this property
of the mass inflation singularity, that will allow us to construct an
approximate solution to the full cross flow and scalar field equations.

 \section{The Continuous Crossflow Model}

The previous idealized models show that when the outflux is concentrated as a
shell, the mass function (and the Weyl curvature invariant) inflates
exponentially.  Originally, this was  shown for a continuous, arbitrary outflux
which starts a finite time after the event horizon. In this section we shall
introduce the notation and Einstein equations for spherical symmetry,  review
the standard mass inflation solution and discuss its limitations.

In a general spherical spacetime it is convenient to introduce a coordinate
system $x^\alpha = (x^a, \theta,\phi)$,  $(a=1,2)$,  where $x^a$ are the
coordinates of the radial two-spaces $(\theta,\phi)= \hbox{constant}$.
Introducing a function $r(x^a)$ which measures the area of the two-spheres, the
metric can be written:
\begin{equation}\label{11}
ds^2 = g_{ab} dx^adx^b + r^2 d\Omega^2,
\end{equation}
where $d\Omega^2$ is the line element of the unit two-sphere and $g_{ab}$ is
the metric of the radial two-spaces.

Scalar fields $m(x^a), f(x^a)$ and $\kappa(x^a)$ can be defined by
\begin{eqnarray}\label{12}
f&=& g^{ab}\partial_a r\partial_b r = 1 - {2m\over r} + {e^2\over r^2},
\nonumber
\\
\kappa &=& -\frac12 \partial_r f = -{1\over r^2}(m - {e^2\over r}).
\end{eqnarray}
The Einstein field equations are then written as
\begin{equation}
G_{\alpha\beta} = 8\pi (E_{\alpha\beta} + T_{\alpha\beta}),
\end{equation}
where $E_{\alpha\beta}$ is the Maxwellian contribution to the stress tensor.
For a   point charge of strength $e$ located at the origin, this  is
\begin{equation}
E_{\alpha}^{\beta} = {e^2\over 8\pi r^4} \hbox{diag}(-1,-1,1,1).
\end{equation}
(We shall assume that $e$ is fixed.)
The non-Maxwellian contribution will be decomposed as
\begin{equation}
^4T^a_b = T^a_b ,\ \  T^\theta_\theta = T^\phi_\phi = P
\end{equation}
where $P$ is the tangential pressure.  The energy-momentum conservation
equation is
\begin{equation}\label{16}
(r^2 T^{ab})_{;b} = (r^2)^{;a} P.
\end{equation}
The field equations can  then be written as
\begin{eqnarray}\label{17}
r_{;ab} + \kappa g_{ab} &=& -4\pi r(T_{ab} - g_{ab} T) \ , \  T =
T^a_a\nonumber \\
R - 2\partial_r \kappa &=& 8\pi (T - 2P).
\end{eqnarray}
Equations (\ref{12}) and (\ref{17}) yield the equations
\begin{equation}\label{18}
m_{,a} = 4\pi r^2(T_a^b - \delta_a^b T) r_{,b}
\end{equation}
which can be used to derive a wave equation for the mass function
\begin{equation}\label{19}
{\ \lower0.9pt\vbox{\hrule \hbox{\vrule height 0.2 cm \hskip 0.2 cm \vrule
height 0.2 cm}\hrule}\ }
 m = 4\pi(2rf(P-T) + \kappa r^2 T) - (4\pi)^2 r^3 T_{ab}T^{ab}.
\end{equation}

In order to be more specific, we will write the two-metric using null
coordinates $U,V$
\begin{equation}\label{20}
g_{ab} dx^adx^b = -2 e^{2\sigma} dU dV.
\end{equation}
We will  choose $V,U$ to be Kruskal coordinates in which  the  asymptotic \RN
metric (mass $m_0$ and charge $e$)  is
regular on the inner horizon. They are related to the usual
Eddington-Finkelstein advanced  and retarded times $v,u$ by
\begin{equation}
U = -e^{-\kappa_0u} ,\ \  V = - e^{-\kappa_0 v}
\end{equation}
where $\kappa_0 = \kappa(r_0)$ is the surface gravity of the  Cauchy horizon
and $r_0$ is its radius.
The wave operator acting on a scalar  $\psi$ takes the form
\begin{equation}
{\ \lower0.9pt\vbox{\hrule \hbox{\vrule height 0.2 cm \hskip 0.2 cm \vrule
height 0.2 cm}\hrule}\ }
\psi = -2 e^{-2\sigma} \psi_{,UV} .
\end{equation}
In this coordinate system, the Ricci scalar is $R = - 2
{\ \lower0.9pt\vbox{\hrule \hbox{\vrule height 0.2 cm \hskip 0.2 cm \vrule
height 0.2 cm}\hrule}\ } \sigma$.

Null radial vectors pointing inwards and outwards can be defined
\begin{equation}
l_a = - \partial_aV ,\ \  n_a = - \partial_a U.
\end{equation}
In the original mass inflation analysis \cite{Poisson}, a null crossflow stress
tensor was used to model the gravitational radiation. This effective
stress-energy model \cite{Isaacson} is justified by the high blueshift near the
Cauchy horizon.  The stress tensor for null crossflowing radiation can then be
written as
\begin{equation}\label{24}
T_{ab} = {L_{in}(V)\over 4\pi r^2} l_al_b +  {L_{out}(U)\over 4\pi r^2} n_an_b
\end{equation}
which satisfies the conservation equations (\ref{16}) and has $P=T = 0$. The
conservation equations  force $L_{in}$ ($L_{out}$) to be a function only of $V$
($U$).

In the Kruskal coordinate, $V$, the  Price power-law tail has the form
\begin{equation}
L_{in}(V) = {dm_{in} \over dv} ({dv\over dV})^2= {\beta \over (-\kappa_0 V)^2}
\left(-\ln(-V)\right)^{-p}.
\end{equation}
As the Cauchy horizon is approached, in the limit, $V \rightarrow 0_-$,
$L_{in}$ diverges and  the source term in the wave equation for $m$ diverges as
well.  The integral solution for the mass function is  \cite{Poisson}
\begin{eqnarray}
\lefteqn{m(U,V) = }  \nonumber \\
 & & \int_{U_1}^U \int_{V_1}^V r'^{-1}e^{-2\sigma'} L_{in}(V') L_{out}(U')
dU'dV' \nonumber \\
& & + m_{in}(V) + m_{out}(U) - m_1.
\end{eqnarray}
The gravitational wave tail influx is turned on at advanced time $V_1$ and the
outflux is assumed to be switched on at the advanced time $U_1$ which is behind
the event horizon.
The divergence of $  L_{in}(V')dV'$  leads to mass inflation with the mass
function behaving as $m\sim 1/V$. Of course, this is only true if the
combination $r^{-1}e^{-2\sigma}$ does not go to zero, but this was proved by
Poisson and Israel \cite{Poisson}.

The previous mass inflation analyses   suffer some limitations. In the picture
presented  \cite{Poisson,Ori}
it is always assumed that the outflux is turned on abruptly after some finite
time behind the event horizon. Essentially, this amounts to the assumption that
a null portion of the Cauchy horizon exists because the solution before the
outflux begins is the Vaidya solution. This segment's existence depends on the
form of the outflux crossing it. If the outflux at early retarded times
($U\rightarrow -\infty$) is too strong, a spacelike singularity will form. The
effect of an outflux crossing a null ray is described by Raychaudhuri's
equation
\begin{equation}\label{27}
{d^2r\over d\lambda^2} = -4\pi r T_{\lambda \lambda}
\end{equation}
where $\lambda$ is an affine parameter on the null hypersurface and $T_{\lambda
\lambda} = T_{ab} {dx^a\over d\lambda}{dx^b\over d\lambda}$ is the transverse
flux and
${dx^b/ d\lambda}$ is tangent to the generators of the null hypersurface. In
the case of interest, the null hypersurface is the Cauchy horizon and $\lambda
\rightarrow -\infty$ corresponds to its ``meeting" with the event horizon at
$H$ in Fig. 1.

Examination of (\ref{27}) shows that in order for  $r$ to be finite   as
$\lambda \rightarrow -\infty$ ,  $T_{\lambda\lambda}$ must satisfy
\begin{equation} \label{28}
\lambda^2 T_{\lambda \lambda} \rightarrow 0  \ \  \hbox{as}\ \  \lambda
\rightarrow -\infty.
\end{equation}
To test this condition we need  to relate the affine parameter $\lambda$ to the
null coordinate $U$ by
\begin{equation} \label{29}
{dU\over d\lambda} = -g^{UV} = e^{-2\sigma}.
\end{equation}
If $e^{-2\sigma} $ diverges as $V\rightarrow 0$ and $U \rightarrow - \infty$,
then depending on $T_{UU}$, the Cauchy horizon may not survive.

The earlier models of mass inflation  do not address this issue since the
outflux is turned on after the event horizon. The corner region, $V\rightarrow
0, -\infty<U<U_1$ is described by a Vaidya solution.  (The metric function
$e^{-2\sigma}$ is  finite  in this region for Vaidya.) For more general models
which include the corner region, the behaviour of $\sigma$ must be found. To do
this, we need to specify the appropriate initial conditions for $T_{UU}$ which
are physically reasonable.

\section{The Outflux}

A star collapsing through its event horizon provides two sources of outflux.
First, the star shines as it collapses and will irradiate the Cauchy horizon
after the event horizon is passed. While we will not attempt to  describe the
actual form of the the star's radiation, we do know that in a freely falling
frame at the event horizon, the radiation must be measured to be bounded.
Kruskal coordinates for the event horizon, $U_+ = e^{\kappa_+u}$ are
appropriate for freely falling observers. These observers measure $T_{U_+U_+}
<$ constant.  Transforming this  to the Kruskal coordinate $U$ appropriate to
 the inner
horizon  the outflux across $CH$ is
\begin{eqnarray}
(T_{UU})_{star} = T_{U_+U_+}\left({dU_+\over dU}\right)^2 \\
\sim (-U)^{-2(1 +
\kappa_+/\kappa_0)} \ \  , \ \ U \rightarrow -\infty. \nonumber
\end{eqnarray}
As we shall see,   the outflux due to the star has a negligible effect compared
to the backscattering of the incoming radiation.

Consider the evolution of a massless spherically symmetric scalar field
in the black hole interior. The characteristic initial value problem is
completely specified by data given on the   the event horizon. The physical
initial data are determined by the Price power law wave tail $v^{-p}$.

\vspace{1em}
\epsfxsize=7.5cm
\epsffile{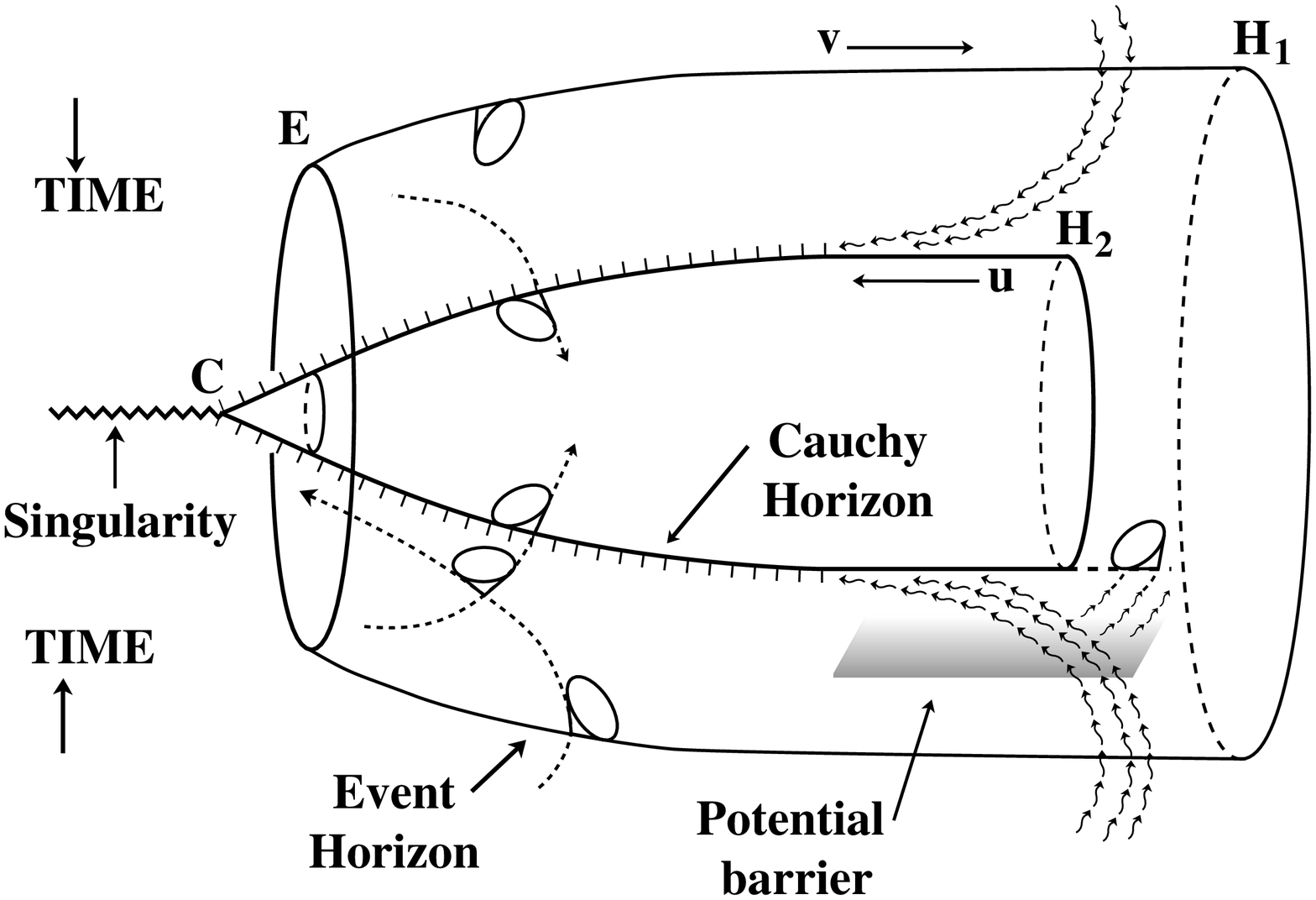}
\begin{center}
\parbox{6.5cm}{
\small
Fig 2. View of the spherical black hole interior (one angular variable
suppressed), showing future light cones and a stream of infalling radiation,
partially scattered off the potential barrier, with the remainder accumulating
along the Cauchy horizon. The point H of Fig. 1. now splits into two points
H$_1$ and H$_2$.
}
\end{center}
\mbox{\hspace{1cm}}

For reference purposes, consider the evolution of a scalar field in a fixed \RN
background with mass $m_0$. The far right hand side of Fig. 2 (with all fluxes
turned off), describes the static \RN solution. It is distinguished by an outer
layer where the gravitational potential barrier is weak and perturbations can
propagate without impediment. The potential  is  peaked around the radius
$r=e^2/m_0$.  This is where most of the perturbation will be scattered. Much
further in, near $r\sim r_0$, the Cauchy horizon is approached and infalling
radiation is strongly blueshifted. It is  important to note that the radiation
is scattered long before it reaches the large blueshift zone.  This motivates
our treatment of the evolution of fields as a scattering problem on a static
\RN background.

Mathematically, scattering of a massless field is given by
${\ \lower0.9pt\vbox{\hrule \hbox{\vrule height 0.2 cm \hskip 0.2 cm \vrule
height 0.2 cm}\hrule}\ } \Psi = 0$, which using the usual advanced/retarded
coordinates is
\begin{eqnarray}
\phi_{uv}&=& V(r) \phi \  \  , \ \  \Psi = \phi/r \nonumber \\
V(r) &=& {f_s(r)\over 4r }  {d \over dr} f_s(r)
\end{eqnarray}
where $f$ has been defined in (\ref{12})  and the subscript $s$ denotes the
static \RN case with mass $m_0$.

This is  a one-dimensional  scattering problem. It is greatly simplified by the
fact that the potential $V(r)$ is highly localized near $e^2/m_0$. It falls off
exponentially \cite{Chandra} in the tortoise coordinate defined by $dx =
dr/f_s(r)$ near the event horizon $x=-\infty$ and the Cauchy horizon $x=
\infty$.
A scalar field will propagate freely near the event and Cauchy horizons and
will only strongly interact with the curvature in the thin belt around
$e^2/m_0$. At the horizons the scalar field solutions will be of the form of
ingoing and outgoing waves $e^{-i\omega v}$ and $e^{-i\omega u}$. The effect of
the potential will be to alter the amplitudes by the reflection and
transmission coefficients, $r(\omega)$ and $t(\omega)$.

If the initial value  on the event horizon is $\phi_0(v)$ then its Fourier
transform  \cite{Chandra}
\begin{equation}
\tilde{\phi_0}(\omega) = {1 \over \sqrt{2 \pi}} \int_{-\infty}^{\infty}
\phi_0(v) e^{i\omega v} dv
\end{equation}
allows us to write the form of the scattered waves as $X(v) + Y(u)$ where
\begin{eqnarray} \label{33}
X(v) &=& {1 \over \sqrt{2 \pi}} \int_{-\infty}^{\infty}
\tilde{\phi_0}(\omega)t(\omega)e^{-i\omega v} d\omega \nonumber \\
Y(u) &=& {1 \over \sqrt{2 \pi}} \int_{-\infty}^{\infty}
\tilde{\phi_0}(\omega)r(\omega)e^{i\omega u} d\omega  .
\end{eqnarray}
The initial conditions are $\phi_0(v) = (\kappa_0 v)^{-p/2}\Theta(v-v_1)$ where
the influx is assumed to start after $v_1$. The Fourier transform behaves
as \cite{Chandra}
\begin{equation}
\tilde{\phi_0}(\omega) \sim\omega^{p/2 - 1}.
\end{equation}
This can be used to calculate the transmitted and reflected flux. The
stationary phase approximation can be used to evaluate the integrals
(\ref{33}). For large $v$ the transmitted flux has the form
\begin{eqnarray}\label{35}
X(v) &\sim& t(\omega_0) (k_0 v)^{-p/2} , \nonumber \\
   \omega_0 = -i p/2v \\
t(\omega_0)  &\sim& 1/v \nonumber
\end{eqnarray}
and the reflected outflux is for large negative $u$
\begin{eqnarray}\label{36}
Y(u) &\sim& r(\omega_0) (-\kappa_0 u)^{-p/2} , \nonumber\\
\omega_0 = -ip/2u \\
r(\omega_0) &\sim& \hbox{constant}. \nonumber
\end{eqnarray}
 We calculated the reflection and transmission coefficients shown in (\ref{35})
and (\ref{36}) using a simple model for the scattering potential: a rectangular
well adjacent to a rectangular barrier. Since the actual scattering potential
looks approximately  like a well adjacent to a sharp-edged barrier,  this model
captures the essential features of the potential.
For low energy scattering it is expected that the perturbations will be
strongly influenced by the potential so that there will be an almost total
reflection.  Actually, it is more appropriate to use the term refraction here
since the reflected beam continues on to smaller radii.

The general effect of the \RN curvature is to scatter the influx $T_{vv} \sim
(\kappa_0 v)^{-p}$ into a reflected flux $T_{uu} \sim \alpha \  (-\kappa_0
u)^{-p}$ and a transmitted flux $T_{vv} \sim \beta (\kappa_0 v)^{-p-2}$ near
the Cauchy horizon, where $\alpha$ and $\beta$ are the reflection and
transmission coeffients and are  $O(1)$.
Is this linearized scattering theory useful for our problem?

Consider the initial layers just beneath the event horizon $r=r_+ - \epsilon$.
This region is far above the strongly blueshifted region, so the flux of energy
is only that of the Price gravitational wave tail $v^{-p}$. For late times this
is very weak and will only be a small perturbation from the vacuum \RN
solution. The effect of the \RN geometry on the wave tail influx will be
negligible and the influx from the event horizon will freely propagate to the
\RN potential barrier.

The black hole interior can be approximated from the event horizon to the
scattering potential as   \RN. Our method will be to solve the Einstein
equations in the interior region after the scattering potential.  Initial
conditions can be set just after the potential barrier, given by the \RN
scattering problem.  In fact, just after the potential barrier the fluxes are
not particularly large since the  blueshift  region has not been approached
yet. Until the Cauchy horizon is reached the perturbations to \RN are small.
Our approach is to find an approximate solution which is good close to the
Cauchy horizon (where perturbations are large) and which satisfies the initial
conditions given by scattering from the potential barrier.

As a first step toward the analytic scalar field approximation, we shall
first introduce a null cross flow solution which incorporates the boundary
conditions discussed here.

\section{Analytic Approximation for light\-like Crossflow}

Our aim is to construct an analytic model for the black hole interior after the
potential barrier. We shall first start with a null crossflow stress tensor and
introduce some approximations. This can then be used as a model for what we
expect to happen in the scalar field evolution.

The null crossflow stress tensor is of the form of equation (\ref{24}) with the
luminosity functions given by scattering
\begin{eqnarray}
L_{in}(V) &=& {\beta (-\kappa_0 V)^{-2}}\left(-\ln(-V)\right)^{-q} \nonumber \\
L_{out}(U) &=& {\alpha (-\kappa_0 U)^{-2}} (\ln(-U))^{-p}
\end{eqnarray}
where $\alpha$ and $\beta$ are dimensionless positive numbers, corresponding to
the reflection and transmission coefficients respectively and $q=p+2$.

Introduce the functions $A(U), B(V)$ defined by
\begin{eqnarray}
L_{out}(U) &=& A''(U) \nonumber \\
L_{in}(V) &=& B''(V)
\end{eqnarray}
where $'$ denotes ordinary differentiation. For \mbox{$V\rightarrow 0$}
\begin{eqnarray}\label{39}
B(V) &=& - {\beta \over \kappa_0^2 (q-1)} \left(-\ln(-V)\right)^{-q +
1}\nonumber \\
& &
\left(1 + {q-1 \over -\ln(-V)} + ...\right)  \nonumber \\
B'(V) &=&   {\beta \over \kappa_0^2 (-V)}\left(-\ln(-V)\right)^{-q  } \nonumber
\\
& &  \left(1 + {q  \over -\ln(-V)} + ...\right)
\end{eqnarray}
and for $U\rightarrow -\infty$
\begin{eqnarray}\label{40}
A(U) &=&   {\alpha \over \kappa_0^2 (p-1)} (\ln(-U))^{-p + 1} \nonumber \\
& &  \left(1 - {p-1
\over  \ln(-U)} + ...\right)  \nonumber \\
A'(U) &=&   {\alpha \over \kappa_0^2 (-U)} (\ln(-U))^{-p  } \nonumber \\
& &  \left(1 - {p  \over \ln(-U)} + ...\right) .
\end{eqnarray}
In the corner region $V\rightarrow 0, U\rightarrow -\infty$, the functions $A$
and $B$ are small, but derivatives of $B$ with respect to $V$   diverge.

We wish to concentrate on the region after the potential barrier at early
times.  Before the potential barrier we expect \RN to be a good model.
 In the innermost regions we must model the effect of the infinite blueshift of
the inflowing radiation.

Using the metric (\ref{20}) we  note that the Einstein equations allow us to
write wave equations for two combinations of the metric functions, which  do
not include the mass function as a source term:
\begin{eqnarray}\label{41}
(\ln(r^{-1}e^{-2\sigma}))_{,UV}& =& - {e^{2\sigma} \over 2 r^2}\left(1 -
{3e^2\over r^2}\right) \nonumber \\
(r^2)_{,UV} &=& - e^{2\sigma}\left(1 - {e^2\over r^2}\right).
\end{eqnarray}

In order to solve the evolutionary problem, we need to find the solution in the
intermediate region after the potential barrier and before the region where $r$
goes to zero. This region will be defined by $r\neq 0$.
When this stipulation is made it is impossible for $r^{-1}e^{-2\sigma}$ to go
to zero \cite{Poisson}. This means that both wave equations (\ref{41}) do not
have any potentially diverging source terms and both will have finite
solutions. For conciseness, introduce the bounded and non-zero variables
\begin{equation}
\chi = r^{-1} e^{-2\sigma} \ ,\  \rho = \frac12 r^2.
\end{equation}

The Einstein equations can then be written as equations (\ref{41}) and the null
hypersurface constraint equations:
\begin{eqnarray}\label{43}
\partial_U(\chi \rho_{,U}) &=& - \chi A'' \nonumber \\
\partial_V(\chi \rho_{,V}) &=& - \chi B''.
\end{eqnarray}
The mass function obeys the wave equation
\begin{equation}\label{44}
m_{,UV} = \chi A''B''.
\end{equation}

As long as $r\neq 0$, we can write a solution with $\chi$ and $\rho$ being
close to their \RN values plus perturbations which are small in this region.
The metric functions for static \RN with a mass $m_0$ will be denoted with a
subscript ``s", so that $f_s(r_s)$ and $\kappa_s(r_s)$ are defined by equation
(\ref{12}). The functions $\rho_s$ and $\chi_s$ and their derivatives are
\begin{eqnarray}\label{45}
\rho_s = \frac12 r^2_s \nonumber \\
\rho_{s,U} = - \frac12 {r_s f_s \over \kappa_0 U} \nonumber \\
\rho_{s,V} = - \frac12 {r_s f_s \over \kappa_0 V}  \\
\chi_s = {-2 UV \over f_s} {k_0^2 \over r_s} \nonumber
\end{eqnarray}
\begin{eqnarray}\label{46}
\chi_{s,U} = -{\kappa_0 V\over r_s^2} \left(1 + {2\kappa_0 r_s \over f_s}
(\kappa_0 - \kappa_s)\right) \nonumber \\
\chi_{s,V} = -{\kappa_0 U\over r_s^2} \left(1 + {2\kappa_0 r_s \over f_s}
(\kappa_0 - \kappa_s)\right).
\end{eqnarray}
In the limit of the Cauchy horizon, ($ UV \rightarrow 0, r_s\rightarrow r_0$)
these functions take on the limiting value
\begin{equation}\label{47}
f_s \rightarrow  -2UV
\end{equation}
\begin{equation}\label{48}
\rho_s \rightarrow  \frac12 r_0^2 ,\
\rho_{s,U} \rightarrow  {r_0V\over \kappa_0} ,\
\rho_{s,V} \rightarrow  {r_0U\over \kappa_0}
\end{equation}
\begin{equation}\label{49}
\chi_s \rightarrow  {\kappa_0^2 \over r_0},\
\chi_{s,U} \rightarrow  -{\kappa_0 V \over r_0^2} ,\
\chi_{s,V} \rightarrow  -{\kappa_0 U\over r_0^2}  .
\end{equation}

We can now construct a solution to the Einstein equations using an iterative
approach, taking the static \RN solution as the zeroth order solution
($\chi^{(0)} = \chi_s, \rho^{(0)} = \rho_s$) and substituting back into the
Einstein equations to find the first order correction terms.  Equations
(\ref{43}) can be integrated to solve for $\rho$:
\begin{eqnarray}\label{50}
\lefteqn{\rho = \rho_s} \\&  - &\int^{V} {dV'' \over \chi(U'',V'')} \int^{V''}
dV'
\chi(U',V') B''(V')  \nonumber \\
&  -&  \int^{U} {dU'' \over \chi(U'',V'')} \int^{U''} dU' \chi(U',V') A''(U').
\nonumber
\end{eqnarray}
It is clear in our approximation scheme that (\ref{50}) is the leading order
contribution to the solution of the Einstein equations. The contribution from
(\ref{41}) will be of lower order.

Integration of (\ref{50}) by parts gives the solution
\begin{equation}\label{51}
\rho  = \rho_s - (A+B) + \epsilon,
\end{equation}
where $\epsilon$ is
\begin{eqnarray}\label{52}
\lefteqn{\epsilon = } \nonumber \\
& &  \int^{V} {dV''  \over \chi_s(U'',V'')}  \int^{V''} dV'
\chi_{s,V'}(U',V') B'(V') \nonumber \\
	   &  &\mbox{\hspace*{-18pt}} + \int^{U} {dU''  \over \chi_s(U'',V'')}
\int^{U''} dV'
\chi_{s,U'}(U',V') A'(U')\nonumber \\
& & \sim  U \int B dV + V \int A dU
\end{eqnarray}
which is much smaller than  $A+B$   in the remote past of CH.
Using this approximation in the second equation of (\ref{41}) and expanding to
first order in $A$ and $B$, allows the estimation
\begin{equation}
\chi = \chi_s + O({1\over r_s^2} (A+B)).
\end{equation}
Substitution of this order of correction back into (\ref{50}) yields a second
order approximation
\begin{eqnarray}
\rho = \rho_s & - & A (1 + O(\ln(-U))^{-p})  \nonumber \\
& - & B (1 +O\left(-\ln(-V)\right)^{-q}).
\end{eqnarray}
 Clearly, in the corner region where $(\ln(-U))^{-1} \sim
\left(-\ln(-V)\right)^{-1}\sim 0$, $\rho$ is well approximated by the leading
order solution (\ref{51}).

To linear order in $A$ and $B$, the mass function can be integrated from
(\ref{44})
\begin{equation}
m(U,V) = \chi_s  A'B' (1 + O(A+B)) + m_{in}' + m_{out}' - m_0
\end{equation}
which in the limit $V\rightarrow 0$ is
\begin{equation}
m \sim {\alpha \beta \over r_0 \kappa_0^2} {1\over UV} (\ln(-U))^{-p}
\left(-\ln(-V)\right)^{-q}
\end{equation}
showing the usual $1/V$ inflation found in earlier work \cite{Poisson}.

The solutions for the original metric functions $r$ and $\sigma$ are
\begin{eqnarray}
r &=& r_s- {(A+B)\over r_s} + {2AB \over r_s^3} + O(A^2 + B^2)\nonumber \\
\sigma &=& \sigma_s + {(A+B)\over2r_s^2} + {AB \over r_s^4} + O(A^2 + B^2).
\end{eqnarray}

This approximation is not so good as the scattering surface is approached ($UV
\rightarrow 1$, so that correction terms (\ref{52}) are comparable to the first
order terms in (\ref{51})).  We already know that the solution near the
scattering surface should be approximately described by the \RN solution.  It
is only after this region, deep into the blueshift region that an approximate
solution is needed and this is where it is important that the solution be
accurate. The solution that we have found is accurate where it matters, close
to the Cauchy horizon.

\section{The Scalar Field Solution}

Using   approximations similar to those just discussed  for lightlike
crossflow, we
can develop an approximate analytic solution for the scalar field equations.
As before the physics tells us that the interior solution can be approximated
well by the static \RN solution from the event horizon, down until the
scattering surface.

The Einstein equations for coupling to a massless scalar field are
\begin{eqnarray}\label{58}
 r\phi_{,UV}&=& - r_{,U}\phi_{,V} - r_{,V} \phi_{,U}  \\
\rho_{,UV} &=&  - 2 {1\over r\chi}\left(1 - {e^2\over r^2}\right) \nonumber \\
(\ln(\chi))_{,UV}& =& 8\pi \phi_{,U}\phi_{,V} - {1\over r^3\chi} \left(1 -
{3e^2\over r^2}\right) \nonumber \\
(\chi \rho_{,U})_{,U}&=& - 8\pi \rho \chi \phi_{,U}^2 \nonumber \\
(\chi \rho_{,V})_{,V} &=& - 8\pi \rho \chi \phi_{,V}^2 \nonumber \\
m_{,UV} &=&  16 \pi^2 \chi r^4  \phi_{,U}^2 \phi_{,V}^2  - 4 \pi r f
\phi_{,U}\phi_{,V}. \nonumber
\end{eqnarray}

Define functions $a(U)$ and $b(V)$ by setting their derivatives equal to
\begin{equation}
a'(U) \equiv \phi_{,U}|_b ,\ \  b'(V) \equiv \phi_{,V}|_b
\end{equation}
where the subscript $b$ refers to the value of the scalar field  given by
scattering at the underside of the  potential barrier.

As before, the wave equations for $\rho$ and $\chi$ have solutions which are
finite and non-zero in the corner region as long as $\phi$ does not diverge.
The initial conditions given by scattering (\ref{35},\ref{36}) are that the
scalar field is initally regular.  Near the scattering surface the radius will
be close to its \RN value, so using the scalar wave equation and the \RN radius
(\ref{45}), it can be seen that the $UV$ mixed derivative of the scalar field,
near the initial surface is
\begin{eqnarray}
\lefteqn{
\phi_{,UV}|_b = \frac12  {f_s\over r_s \kappa_0} ({\phi_{,V} \over U } +
{\phi_{,U} \over V}) |_b} \nonumber \\
&& \sim \left(-\ln(-V)\right)^{-q/2} + (\ln(-U))^{-p/2}.
\end{eqnarray}
This derivative is small in the corner ($V \rightarrow 0, U\rightarrow -\infty
$), so in the earliest regions the scalar field will not be changing rapidly
from its initial value. This motivates us to make the ansatz that the leading
order behaviour of the scalar field, near the Cauchy horizon should be
\begin{eqnarray} \label{61}
\phi_{,U}^{(0)} = a'  &\equiv &\sqrt{A'' \over 4\pi r_0^2} \nonumber \\
&\sim & {1\over
\sqrt{4\pi r_o^2 \kappa_0^2}} {1\over (-U)} (\ln(-U))^{-p/2} \nonumber \\
\phi_{,V}^{(0)} = b' &\equiv& \sqrt{B'' \over 4\pi r_0^2}  \\
&\sim& {1\over
\sqrt{4\pi r_o^2\kappa_0^2}} {1\over (-V)} \left(-\ln(-V)\right)^{-q/2}.
\nonumber
\end{eqnarray}
With this ansatz, we can see that the scalar field will be small everywhere in
the corner region, but that derivatives with respect to $V$ will diverge near
the Cauchy horizon.

As before we can calculate the first order correction terms by iterating the
Einstein equations, again taking the zeroth order solutions for $\rho$ and
$\chi$ to be the same as the \RN solution.  The solution for $\rho$ is the same
as the lightlike cross flow solution (\ref{51}).  Substitution of (\ref{51})
and (\ref{61}) into the scalar wave equation yields the first order equation
\begin{equation}
\phi_{,UV}  =  -{1\over 2 \rho_s}( (\rho_{s,U} - A')b' + (\rho_{s,V} - B')a')
\end{equation}
which can be integrated asymptotically in the corner region, making use of
(\ref{45})
\begin{equation}
\phi   = a + b + {1\over r_0^2}(Ab + Ba) + O( UV (a+ b)).
\end{equation}
Using the first order solutions for $\phi$ and $\rho$ the wave equation for
$\chi$ can be integrated
\begin{equation}
\ln\chi  = \ln\chi_s + 8 \pi ab + O(A+B).
\end{equation}
To leading order the mass function is integrated to be
\begin{equation}
m(U,V) \sim \kappa_0^2/r_0 A'B'
\end{equation}
and the metric functions $r$ and $\sigma$ are
\begin{eqnarray} \label{66}
r &=& r_s - {(A+B)\over r_s} - {AB \over r_s^3} + O(A^2 + B^2)\nonumber \\
\sigma &=& \sigma_s -4\pi ab+ {(A+B)\over 2 r_s^2} + {AB \over r_s^4} +  \\
& & O(A^2 + B^2). \nonumber
\end{eqnarray}

The existence of the Cauchy horizon in this solution can now be tested.
Substitution of the solution for $\sigma$ given by (\ref{66})  into (\ref{29})
gives the following asymptotic relation between the affine parameter,
$\lambda$, and the coordinate, $U$,
\begin{equation}
\lambda = {U \over \kappa_0^2} (1 + O( a)) \  , U\rightarrow -\infty .
\end{equation}
Condition (\ref{28}) for the scalar field solution now reads
\begin{eqnarray}
\lefteqn{\lim_{\lambda \rightarrow -\infty} \lambda^2 T_{\lambda \lambda} = }\\
&& \lim_{U \rightarrow -\infty}  {(\ln(-U))^{-p+2} \over 4 \pi r_0^2
\kappa_0^6}
\nonumber \\
&&
\hspace*{1cm}
\left( 1
+ O(\ln(-U))^{-p+1} \right)
 = 0. \nonumber
\end{eqnarray}
Since condition (\ref{28}) is satisfied, the Cauchy horizon exists in our
approximate solution to the Einstein-scalar field equations. This is of course
evident directly from the asymptotic form of the metric given by (\ref{20}) and
(\ref{66})

\section{Conclusions}
We have found an approximate solution to the scalar field equations in the
black hole interior which includes a complete past segment of the Cauchy
horizon
  and contains the requisite number of arbitrary functions to be
considered ``general". In our picture (see Fig. 2) the Cauchy horizon will be
focussed by the outgoing flux, forcing it to eventually contract to $r=0$,
  forming a   singularity which is possibly spacelike. But we find that the
$r=0$
singularity does not precede the Cauchy horizon in the earliest region of the
interior.

This is not in agreement with a previous numerical study \cite{Gnedin}. That
work consisted of two simulations, in the first of which \cite{Gnedin2} initial
conditions were placed on the event horizon. The result of this simulation was
that the Cauchy horizon survived
the introduction of a scalar field. In the second study, initial conditions
were
set outside the event horizon, which in effect allows tails to form. These
results (see their Fig. 4) show that any null portion  is abbreviated or
perhaps completely absent. It is difficult to judge from their evidence, but
the authors state that the Cauchy horizon is destroyed and replaced by a
spacelike $r=0$ singularity. Further numerical work on this problem is in
progress \cite{Choptuik}, which should help clear up the discrepancy in
results.

\section{Acknowledgments}
We would like to thank Pat Brady and Eric Poisson for stimulating discussions.
This work was supported by the Natural Sciences and Engineering Research
Council of Canada. S.M.M. wishes to acknowledge an Avadh Bhatia Fellowship.


\begin{thebibliography}{99}
\bibitem{Wald}R.M. Wald, {\em General Relativity}, (University of Chicago
Press, Chicago, 1984), pp. 322-324.

\bibitem{Price1}R.H. Price, Phys. Rev {\bf D 5}, 2419(1972); {\bf D 5}, 2439
(1972).

\bibitem{BKL}V.A. Belinskii, E.M. Lifshitz and I.M. Khalatnikov, Adv. Phys.
{\bf 19}, 	525 (1970); C.W. Misner, Phys. Rev. Lett. {\bf 22}, 1071
(1969).

\bibitem{Penrose} R. Penrose, in {\em Battelle Rencontres,}  ed.s C.M. De Witt
		and J.A. Wheeler, (W. A. Benjamin, New York, 1968), p. 222.

\bibitem{Ellis}G.F.R. Ellis and A.R. King, Commun. Math. Phys. {\bf 38}, 119
(1974). A.R. King, Phys. Rev. {\bf D 11}, 763 (1975).

\bibitem{Chandra}S. Chandrasekhar and J.B. Hartle, {\em Proc. R. Soc. London}
             	{\bf A284}, 301 (1982); Y. Gursel, V.D. Sandberg, I.D. Novikov,
and A.A. Starobinsky, 		Phys. Rev. {\bf D19}, 413 (1979); Y. Gursel,
I.D.
Novikov, V.D. Sandberg, and A.A. Staro\-bin\-sky, Phys. Rev. {\bf D 20}, 1260
(1979).

\bibitem{Poisson}  E. Poisson and W. Israel, {  Phys. Rev.}
		{\bf D41}, 1796, (1990).

\bibitem{Ori}A. Ori, Phys. Rev. Lett. {\bf 67}, 781 (1991).

\bibitem{Ori2}A. Ori, Phys.  Rev. Lett. {\bf 68}, 2117 (1992); S. Droz, {\em
Lake Louise Winter Institute Proceedings, 1994}, to be published; A. Bonanno,
S. Droz, W. Israel and S.M. Morsink, Can. J. Phys., to be published; P.R.
Brady, S. Droz, W. Israel and S.M. Morsink, paper in preparation.

\bibitem{Yurtsever} U. Yurtsever, Class. Quantum Grav. {\bf 10}, L17 (1993).

\bibitem{Gnedin}M.L. Gnedin and N.Y. Gnedin, Class. Quantum Grav. {\bf 10},
1083 (1993).

\bibitem{Bonanno}A. Bonanno, S. Droz, W. Israel and S.M. Morsink, Phys. Rev.
{\bf D}, to be published.

\bibitem{Gundlach1}C. Gundlach, R.H. Price and J. Pullin, Phys. Rev. {\bf D49},
883 (1994).

\bibitem{Gundlach2}C. Gundlach, R.H. Price and J. Pullin, Phys. Rev. {\bf D49},
890 (1994).

\bibitem{DTR}T. Dray and G. 't Hooft, Commun. Math. Phys. {\bf 99}, 613 (1985).

\bibitem{Barrabes}C. Barrab\`{e}s, W. Israel and E. Poisson, Class. Quantum
Grav. {\bf 7}, L273 (1990).

\bibitem{Anderson}W.G. Anderson, P.R. Brady, W. Israel, and S.M. Morsink, Phys.
Rev. Lett. {\bf 70},  1041 (1993).

\bibitem{Isaacson}R.A. Isaacson, Phys. Rev. {\bf 160}, 1263 (1968); {\bf 160},
1272 (1968).

\bibitem{Gnedin2}N.Y. Gnedin and M.L. Gnedin, Sov. Astron. {\bf 36}, 216,
(1992).

\bibitem{Choptuik}M.W. Choptuik, personal communication; C. Gund\-lach,
personal
communication.
\end{thebibliography}
\end{document}